# Adaptive Approach for Situational Analysis of Space Experiments

Atanas Marinov Atanassov

*Solar Terrestrial Influences Institute, Bulgarian Academy of Sciences, Stara Zagora Department, P.O. Box 73, 6000 Stara Zagora, Bulgaria*

## 1. Introduction

In order to solve scientific and practical problems, connected with satellite experiments, it is necessary to analyze a set of geometrical and physical conditions- situational analysis. It allows to determine proper time intervals for conducting experiments and measurements and for optimizing complex and expensive scientific programs. In the practice of the situational analysis, different restricting conditions are verified for example:

- a satellite crosses the Earth shade [1];
- the angle between the observed and another bright object on the celestial sphere (Sun, Moon, bright stars) shall not to be below a specified value [2];
- a satellite passes over a territory from the Earth surface [3];
- a satellite is in the visibility zone of an observation point of the Earth surface;
- the Earth radiation background is within admissible limits in regards to the conducted measurements [2];
- a satellite is properly orientated towards the force lines of the Earth magnetic field;
- a satellite is in the zone of the Earth radiation belts, in the zone of the shock wave or in the magnetopause of the Earth magnetosphere [1];
- the vision axis of the optical instrument is in a region of the celestial sphere or the Earth surface which is interesting in regards to the conducted experiment;
- the magnetic force line where the measuring instrument is, pierces the polar oval region.

The proper orientation of the model of the space platform with scientific instrumentation with a view to the simulation of experiments and measurements in the objective space will be called a **situation**. Each situation can be presented in the common case by a predicate function S:

$$S = S(\vec{R}, A, t) = \overline{0,1} \qquad (1)$$

In (1) $\{\vec{R}\} = <\vec{r}_1(t), \vec{r}_2(t),...,\vec{r}_n(t),>$ - the multitude of radius-vectors of objects in the model space; $\{A\} = <\alpha_1(t), \alpha_2(t),...,\alpha_m(t)>$ - the multitude of vector or scalar fields, describing definite properties of the model space and t- the time. Due to the complex character of the space processes and phenomena, S can have a complex presentation. Practically, we might have a combination of several restricting conditions which are independent of one another. Thus, the multitude of conditions $\{\gamma_i\}$ can be juxtaposed to the multitude of predicate functions $\{s_i\}$. The realization of situation S requires the realization of identity:

$$S = s_1 \wedge s_2 \wedge ... \wedge s_n = 1 \qquad (2)$$

## 2. Formulation of the Problem

Obviously, in order to verify identity (2) it is necessary to compute the predicate function $s_i$ which is related to considerable computation expenses. With a view to (2), $\{s_i\}$ can be considered arranged in the sense of the computation sequence of $s_i$. When $s_k = 0$, the further computation of the predicate functions is terminated in case that (2) is verified by the scheme:

$$S = (...(s_1 \wedge s_2) \wedge ... \wedge s_{n-1}) \wedge s_n \qquad (3)$$

In (3) each predicate function $s_i$ is computed if the preceding ones in $\{s_i\}$ are equal to one. A measure of ineffectiveness for scheme (3) can be the computation expenses of $s_i$ at $j < k$ when $s_k = 0$.

It is important to have in mind that each $\gamma_i$ is executed within the range of a temporal interval $T_i$ and is not executed for the next $T_i^*$. The only practically obtained requirements to $T_i$ and $T_i^*$, which will be taken into account, are $T_i, T_i^* \gg \Delta t$.

### 3. Adaptation and Optimization Strategy

Problem A will be called inverse to B if the aims of A and B are contrary [6]. In our case the problem for determination of the temporal interval, in which the conditions $\{\gamma_i\}$ are executed, is inverse to the problem for determination of its adjacent one in which they are not executed or, according to (3) at least one of them is not executed. Obviously, this problem is more efficient since it is adequate to check whether only one condition $\gamma_i$ is not executed.

$$\overline{S} = \overline{s_1 \wedge s_2 \wedge ... \wedge s_k \wedge ... \wedge s_n} = \overline{s}_1 \vee \overline{s}_2 \vee ... \vee \overline{s}_k \vee ... \vee \overline{s}_n \qquad (4)$$

In fact, according to the above-said, $\Delta t \ll T_i$. If for the discretization moment t we have $s_k = 0$, then we can assume that for t+$\Delta$t this condition will not be executed again.

The strategy for implementation of situational analysis is reduced to a consecutive conditions verification. If all conditions are fulfilled, then we have a situation of the searched type. In case that a dissatisfied condition is found out for moment t, then there isn't any situation. When for t' this condition is once again dissatisfied, then the next condition in multitude $\{\gamma_i\}$ will be verified.

We are going to examine two approaches, connected with the organization of the $\{\gamma_i\}$ conditions verifications. The first approach is connected with replacement of the first unfeasible condition which is in the first place of the multitude of conditions [5]. Instead of (4) as a result of the disjunction commutation we can write down:

$$\overline{S} = \overline{s}_k \vee \overline{s}_1 \vee \overline{s}_2 \vee ... \vee \overline{s}_n \qquad (5)$$

The verification algorithm for (3) always begins with the first element. The replacement of every unfeasible condition when applying the alternative strategy is equivalent to adaptation of (5) to the model medium conditions.

The second approach requires to treat the multitude of conditions as a ring-shaped structure. This means that the last element of $\{\gamma_i\}$ is followed by the first one.

$$\overline{S} = \overline{s}_k \vee \overline{s}_{k+1} \vee ... \vee \overline{s}_n \vee \overline{s}_1 \vee ... \vee \overline{s}_{k-1} \qquad (5')$$

This allows to apply the alternative strategy without rearranging the multitude of conditions only by moving a pointer along this ring-shaped structure until finding out an unfulfilled condition. A situation is found out upon execution of a full run along the ring-shaped structure.

### 4. Analysis and Effectiveness Estimation.

In order to estimate the effectiveness, the following two cases are of interest:
a). The predicate functions $s_i$ on the basis of which identity (2) is verified which are characterized by equal or almost equal computational expenses $\xi \cong \text{const}$.
b). The predicate functions $s_i$ on the basis of which identity (2) is verified which are characterized by different computational expenses.

In the first case $(k-1).\xi$ computational units will be saved for each performed step $\Delta t$ based on the offered strategy, in contrast to the direct application of (3) where the dissatisfied condition is in position k.

For illustration we present an analysis, connected with the pass of a couple of satellites over the visibility zone of a ground-based observation station, proper mutual configuration and distance between them, lack of factors, hampering the observation of satellites from the Earth (Moon), lack of factors, preventing the observation of the optical instruments from one of the satellites (Moon, Sun). This situation treats the problem for a possible synchronous observation from the Earth surface and from one of the satellites of an emission, caused by neutral particles injection for studying the earth ionosphere (the experiment "Aktiven").

In the second case where the computational expenses for the separate conditions are different, we can write down $\sum_{i=1}^{k-1}\xi_i$ for the saved computational expenses. It is worth mentioning that the arrangement of the conditions in $\{\gamma_i\}$ (3) is significant. Despite the quasi casual choice of an unfeasible condition in applying the alternative strategy, there are certain peculiarities. The first algorithm may involve a condition with large computational expenses to the first position. If this condition remains in the beginning of the multitude of conditions $\{\gamma_i\}$ then the analysis efficiency can be reduced. With the second approach this problem does not exist as a result of the constant arrangement of the conditions and their circulation.

We have an example of a situational problem of the second type when the analysis includes conditions, connected with the parameters of medium $\alpha_i$. The condition for the satellite velocity vector to conclude a proper angle with the magnetic field vector $\vec{B}(\vec{r},t)$ is important in the analysis of active experiments. (The "Aktiven" experiment). The radiation background is important in the astrophysical observations of Roentgen sources [2]. When phenomena connected with particles dissipation in the polar oval region are investigated, it is important to measure the magnetic field in the region of the force lines along which the dissipation occurs. In all presented cases the verified conditions are connected with complex model computations.